\documentclass[aps,prl,preprint,psfig,groupedaddress]{revtex4-1}
\usepackage{graphicx}
\usepackage{float}
\newcommand\mc[1]{\multicolumn{1}{c}{#1}} 
\newcommand\T{\rule{0pt}{2.6ex}}       
\newcommand\B{\rule[-1.2ex]{0pt}{0pt}} 
\DeclareRobustCommand{\rchi}{{\mathpalette\irchi\relax}}
\newcommand{\irchi}[2]{\raisebox{\depth}{$#1\chi$}}

\begin{document}
 \title{Characterization of switching field distributions in Ising-like magnetic arrays}
 \author{Robert D. Fraleigh}
 \affiliation{Department of Physics, The Pennsylvania State University, University Park, Pennsylvania 16802-6300, USA}
 \author{Susan Kempinger}
 \affiliation{Department of Physics, The Pennsylvania State University, University Park, Pennsylvania 16802-6300, USA}
 \author{Paul Lammert}
 \affiliation{Department of Physics, The Pennsylvania State University, University Park, Pennsylvania 16802-6300, USA}
\author{Sheng Zhang}
 \affiliation{Materials Science Division, Argonne National Laboratory, 9700 S. Cass Avenue, Argonne, IL 60439, USA}
 \author{Vincent H. Crespi}
 \affiliation{Department of Physics, The Pennsylvania State University, University Park, Pennsylvania 16802-6300, USA}
 \author{Peter Schiffer}
 \affiliation{Department of Physics and the Frederick Seitz Materials Research Laboratory, University of Illinois at Urbana-Champaign, Urbana, Illinois 61801, USA}
 \author{Nitin Samarth}
 \email{nsamarth@psu.edu}
 \affiliation{Department of Physics, The Pennsylvania State University, University Park, Pennsylvania 16802-6300, USA}

 \date{\today}
 
\begin{abstract}
The switching field distribution within arrays of single-domain ferromagnetic islands incorporates both island-island interactions and quenched disorder in island geometry. Separating these two contributions is important for disentangling the effects of disorder and interactions in the magnetization dynamics of island arrays. Using sub-micron, spatially resolved Kerr imaging in an external magnetic field for islands with perpendicular magnetic anisotropy, we map out the evolution of island arrays during hysteresis loops. Resolving and tracking individual islands across four different lattice types and a range of inter-island spacings, we extract the individual switching fields of every island and thereby determine the relative contributions of interactions and quenched disorder in the arrays. The width of the switching field distribution is well explained by a simple model comprising the sum of an array-independent contribution (interpreted as disorder-induced), and a term proportional to the maximum field the fully polarized array could exert on a single island. We conclude that disorder in these arrays is primarily a single-island property.

\end{abstract}
\maketitle
 
Ordered arrays of nanoscale single-domain ferromagnetic islands provide a well-defined Ising system at a spatial scale where it is possible to resolve every Ising degree of freedom. Geometries with frustrated interactions, such as artificial realizations of ``spin ice,'' are particularly interesting because they allow direct visualization of magnetic frustration in a well-controlled model environment \cite{Wang2006, Qi2008, Nisoli2013}. Recent experiments have used artificial spin ice arrays to study the nature of the frustrated ground state \cite{Ke2008b, Morgan2011, Morgan2013}, the effect of thermal fluctuations \cite{Kapaklis2014}, the emergence of effective magnetic charges\cite{Ladak2011, Mengotti2011}, and disorder \cite{Budrikis2014}. However, a deeper understanding of the relationship between the experimental arrays and theoretical models requires a more precise quantification of the relative strengths of disorder and interactions in the experimental systems. The recent development of ferromagnetic island arrays with perpendicular anisotropy \cite{Mengotti2009,Zhang2012a} provides an important opportunity in this regard, in that these arrays are amenable to polar magneto-optical Kerr effect (MOKE) studies. Kerr imaging can potentially resolve array dynamics at the individual-island level, imaged across an entire array, during field sweeps: individual Ising degrees of freedom can be tracked exhaustively not only in space, but also in time. Furthermore, the pairwise interaction between two perpendicular moments depends only on the separation between them, unlike the more complex anisotropic interactions in systems with in-plane moments.

A variety of methods have been used to investigate the static and dynamic magnetic behavior of both in-plane and perpendicular anisotropy arrays. For example, magnetic force microscopy (MFM) imaging has been used to study how individual islands behave within small bit-patterned media arrays\citep{Li2011} ($\sim$100 islands), using the remanent states and coarse field bins. Thermal fluctuations of in-plane islands have been imaged using X-ray magnetic circular dichroism photoemission electron microscopy (XMCD-PEEM) to resolve the individual islands\citep{Kapaklis2014}. However, only magneto-optical methods can spatially map the evolution of an array's magnetization continuously in an external magnetic field at timescales which allow for a quasi-dynamic exploration of a system's microstates\citep{Cormier2008}. In particular, exhaustive statistical analysis of the switching field distributions of magnetic nano-arrays can precisely quantify the role of static disorder in island reversal dynamics -- this information is critical to understanding the relative roles of island disorder and island-island interaction in the dynamics of artificial spin ice. 

We use high-resolution polar MOKE to isolate and detect the magnetic state of individual islands within arrays in a continuously varying external field. We thereby directly measure the distribution of switching fields \textit{in situ} for arrays of several thousand islands. The width of the switching field distribution (here called $\sigma$, with the dimension of magnetic field) should have contributions from both quenched disorder and dipolar interactions between islands. We thus try the simple model

\begin{equation}
\sigma(L)=\alpha K B_0(L)+\sigma_d. 
\label{SigmaEquation}
\end{equation}
for the width $\sigma (L)$ of the switching field distribution of an array of inter-island spacing $L$. Here, $B_0(r)=\frac{4\pi}{\mu_0}\frac{3\vec{r}(\vec{r}\cdot\vec{m})-\vec{m}}{|\vec{r}|^3}$ is the magnetic field strength of a single point dipole at a distance $r$ and $\sigma_d$ represents quenched disorder. 
 $K$ is an effective coordination number implicitly defined by the condition that $K B_0(L)$ is the field the entire rest of a polarized array would exert on a given island. $K$ depends only on the geometry, and therefore gives sigma a geometry dependence suppressed in the notation.

The effective coordination numbers for hexagonal, kagome, square and triangular lattices are 4.53, 5.52. 5.91, and 7.58 respectively; these values exceed the nearest-neighbor lattice coordination numbers due to the contribution from further neighbors. Here, we approximate the field experienced by a given island by the value at the island center, assuming pure dipolar fields from nearby islands. Finally, the term $\alpha$ is a correction factor: if  the fitting form of Equation \ref{SigmaEquation} is physically well-grounded, then $\alpha$ will be a simple constant of proportionality of value close to 1. Several questions present themselves upon consideration of this scaling form. \textit{A priori,} it is not clear whether the disorder contribution can be entirely identified with effective individual-island properties (e.g. variations in shape and edge roughness) or whether random local variations in lattice geometry may also enter in an essential way. Also, it is not clear \textit{a priori} whether this effective coordination number $K$ fully captures the effect of lattice geometry on the switching field distribution: i.e. is $\alpha$ actually constant across different lattice geometries? Since  $\alpha K B_0(r) \propto r^{-3}$ and $\alpha$ is the only component that cannot be calculated from other physical properites, functional fits to Eqn. \ref{SigmaEquation} use the form $a r^{-3}+\sigma_d$, and $\alpha$ is then extracted from the parameter $a$.

\begin{figure}[h]
\includegraphics[scale=0.75]{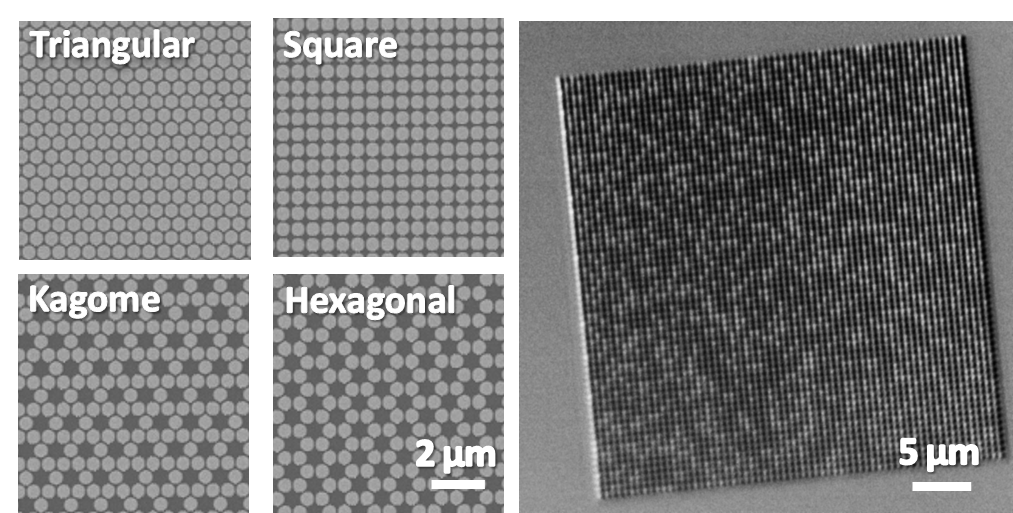}
\caption{Left: Scanning electron microscopy images of arrays with different geometries. Partial arrays are shown. Full arrays measure $35\,\mu m\times 35\,\mu m$. All images shown are of arrays with a inter-island spacing of 500 nm. Right: MOKE contrast in a 500 nm inter-island spacing square array, near the coercive field during a hysteresis loop.}
\label{SEM}
\end{figure}

Electron beam (e-beam) lithography was used to define $450\,$nm diameter circular islands in both non-frustrated (square and hexagonal) and frustrated (kagome and triangular) geometries, using standard lift-off of a bilayer PMMA/PMGI resist stack. Magnetostatic dipolar interaction strength was tuned by varying the array inter-island spacing. Pt/Co multilayer stacks in the sequence Ti($2\,$nm)/Pt($10\,$nm)/[Co($0.3\,$nm)/Pt($1\,$nm)]$_{8}$ were deposited using DC sputtering; such multilayers have strong perpendicular anisotropy and nearly square hysteresis loops\citep{Lin1991}. Bulk properties were measured using superconducting quantum interference device magnetometry (SQUID). Hysteresis loops up to $\pm5$ Tesla were measured for fields both parallel and perpendicular to the sample plane. The easy axis is perpendicular to the plane, confirmed by sharp, nearly square hysteresis loops with a coercivity of $350\,$G, $M_s=375\times 10^3$ A/m and anisotropy constant $K_1 = 94 \times 10^3$ J/$\mathrm{m}^3$. Scanning electron microscopy shows that the individual nanomagnets in the array have an edge roughness of $\sim$5 nm and a corresponding diameter variation of approximately 5 -- 10 nm. Atomic force microscopy shows a surface roughness of $\sim$1 nm and a height of 22 -- 23 nm, which is consistent with the nominal film thickness of 22.4 nm. This height includes the buffer layer, which has a nominal thickness of 12 nm, so we take the thickness of the magnetically active region to be 10.4 nm. We fabricated arrays with nearest-neighbor center-to-center spacings ranging from 500 nm to 800 nm. Individual islands are well-separated even at the smallest spacing, as shown in Figure \ref{SEM}. The chip contains four different arrays for each combination of lattice geometry and inter-island spacing.

We measured the switching field distributions (SFDs) using polar MOKE.  Our optical setup is an optimized Kerr reflectometry system to allow for diffraction-limited spatial imaging. The maximum spatial resolution of a Kerr microscopy system is described by the Abbe diffraction limit, $d = \frac{\lambda}{2n\sin{\theta}}$, where $n\sin{\theta}$ is the numerical aperture (N.A.) of the objective lens and $\lambda$ is the wavelength of light. We use a 100x oil objective lens (1.3 N.A.) to attain diffraction-limited spatial resolution (150 nm -- 300 nm) using white light filtered to the visible spectrum (400 nm -- 700 nm), which lets us clearly resolve the 450 nm diameter nanomagnets in each array. Island arrays were fabricated to be approximately $35\,\mu m\times 35\,\mu m$ so that the entire array fits within the $35\,\mu m\times 50\,\mu m$ field of view of the Kerr imaging setup. Lateral drift of the array is inevitable for field sweeps lasting several tens of minutes. The centroid for the collection of pixels associated with each island was isolated and tracked throughout a saturating magnetic field sweep. Field sweeps range from ${-800}$ to $800\:\mathrm{G}$. In the switching region from $150$ to $500\:\mathrm{G}$, we use $2\:\mathrm{G}$ steps, and outside this region, we use $40\:\mathrm{G}$ steps.  \par
\begin{figure}[t]
	\includegraphics[scale=0.65]{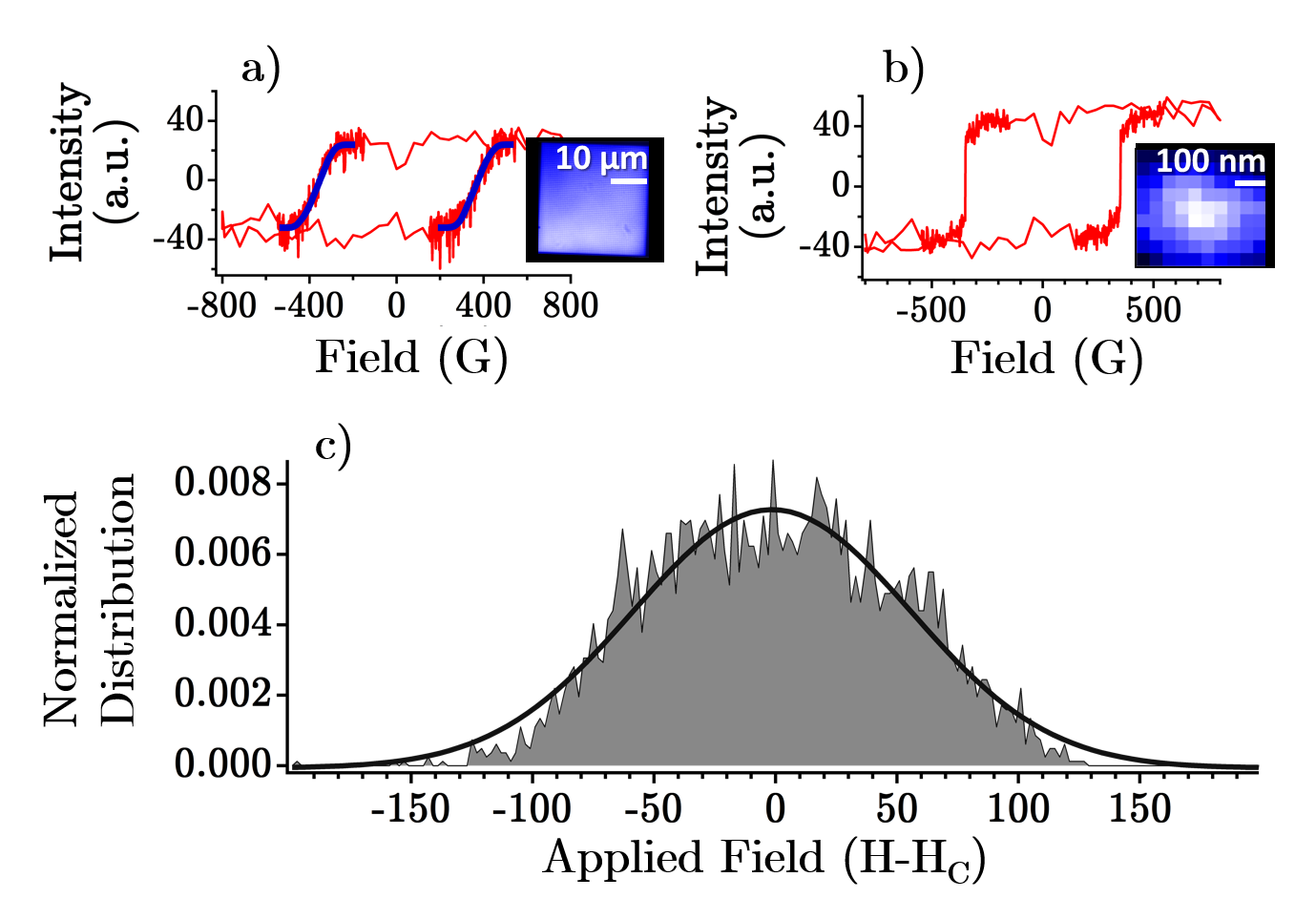}
	\caption{(a) Raw hysteresis loop calculated across an entire array (red) compared to the refined hysteresis loop calculated by combining contributions from individually resolved island switching fields (blue). The inset is a MOKE image of the entire array. (b) Hysteresis loop of an individual island and its pixelated image. 
(c) Histogram of switching fields averaged over several runs, with Gaussian fit. The histogram is centered at the coercive field $H_c$. All data in this figure were taken on a 500 nm square array.}
	\label{Distributions}
\end{figure}

 
By the Kerr effect, island reversal manifests as a fractional change in island intensity, linear in magnetization. The representative image in $\mathrm{Fig.}\:$\ref{SEM} shows the MOKE contrast between islands; $\mathrm{Fig.}\:$\ref{Distributions} provides a raw hysteresis field sweep and the hysteresis loop of an individual island. The sharp switching behavior of an individual island agrees with MFM measurements from previous studies that show that islands of this diameter  have a single domain\citep{Zhang2012a}. The switching behavior has been studied elsewhere for similar magnetic structures\citep{Gerrits2002}, showing coherent domain rotation on the order of several tens of picoseconds, much faster than the acquisition time for a frame in this study and consistent with our observation that an individual island transitions between magnetic states in a single field step. Full-array hysteresis loops can be calculated in two ways: unresolved ``raw'' hysteresis loops that average the MOKE intensity across the entire image and fully resolved ``refined'' loops that enumerate the actual individual islands that switch at any given field step. These two methods agree closely, as shown in $\mathrm{Fig.}\:$\ref{Distributions}a.

\begin{figure}
\includegraphics[scale=0.65]{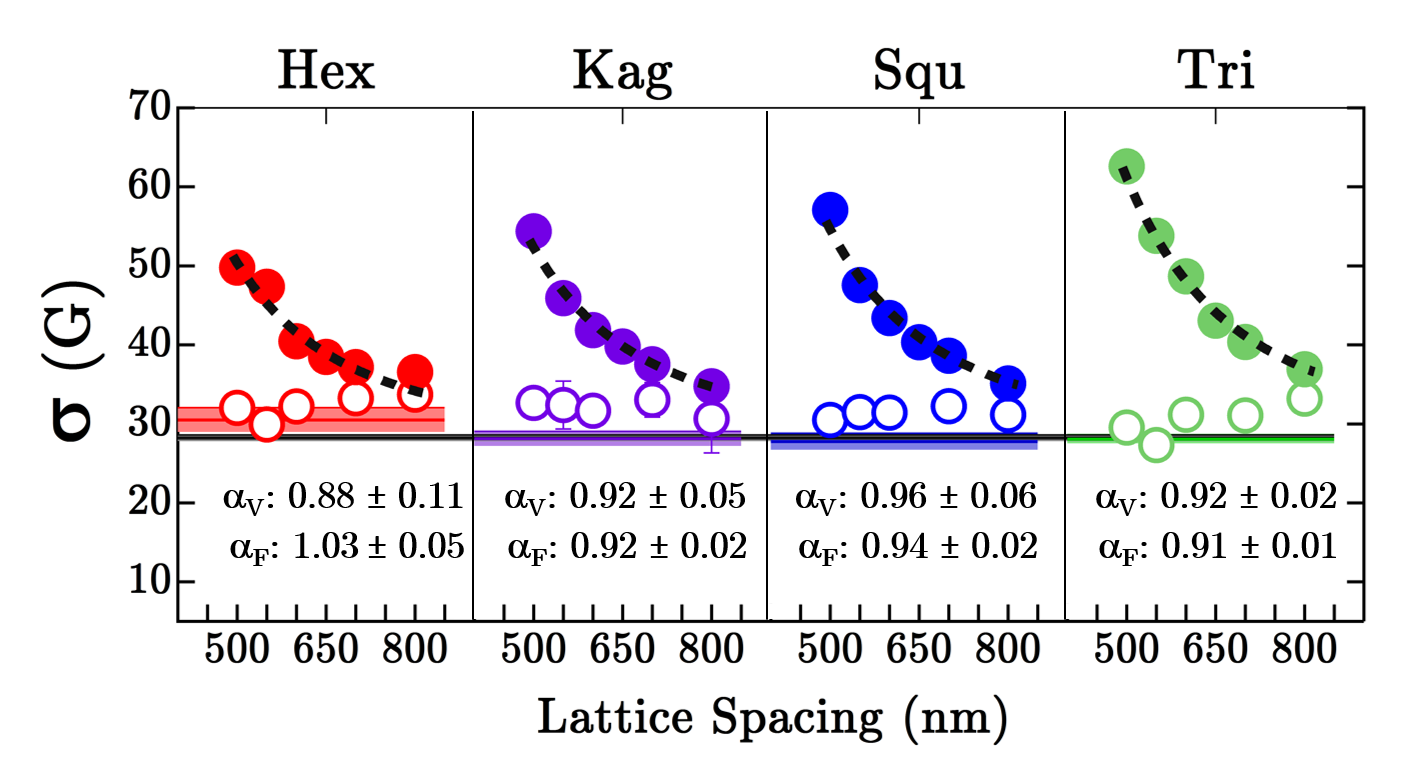}
\caption{(a) Inter-island spacing dependent values of $\sigma$ (closed circles) along with the ``intrinsic'' values of sigma from the $\Delta H$ method (open circles). The value of $\sigma_d$ from fitting to $\mathrm{Eqn.}\:$ \ref{SigmaEquation} for each array type is shown in each panel as a thick horizontal colored line, along with the global average value shown as a thin black line extending across the entire width of the figure. The fits to $\mathrm{Eqn.}\:$\ref{SigmaEquation} with $\sigma_d$ held to its global average are shown as black dashed lines. The fitted values of $\alpha$ for both variable ($\alpha_V$) and fixed $\sigma_d$ ($\alpha_F$) are shown as numerical values. 
}
\label{SigmaData}
\end{figure}

Using the switching fields of individual islands in the array, we directly measure the distribution of average switching fields from several hysteresis loops and fit the results to a Gaussian distribution ($\mathrm{Fig.}\:$\ref{Distributions}c) to extract the width $\sigma$, a global property of the array. Figure \ref{SigmaData} shows values of $\sigma$ for each lattice examined and the corresponding fits to Equation \ref{SigmaEquation}. Fits were carried out using the Levenberg-Marquardt algorithm, and have reduced $\rchi^2$ values of 7.53, 2.54, 2.28, and 0.71 with 5 degrees of freedom for hexagonal, kagome, square, and triangular arrays respectively.

The initial curve fitting finds a value of $\sigma_d$ for each lattice type; however, since $\sigma_d$ arises from physical properties of an island and all islands are fabricated simultaneously, it is reasonable to assume $\sigma_d$ is a constant across all lattices on the same chip. The calculated values of $\sigma_d$ from the curve fits are consistent with this assumption. To treat $\sigma_d$ as a global property of all arrays, we average the values of $\sigma_d$ obtained from fits to each lattice type and then recalculate each fit with $\sigma_d$ fixed to this average value. The results are shown in $\mathrm{Fig.}\:$\ref{SigmaData}.

To quantify how well the global width in switching field is described by the local information in Eqn \ref{SigmaEquation}, we first turn to the $\Delta$H(M,$\Delta$M) method\cite{Berger2005} used in bit-patterned media to extract the so-called ``intrinsic'' portion of the switching field distribution. The intrinsic portion solely arises from the properties of the individual islands and disregards any contributions from interactions. This method involves subtracting a series of first-order reversal curves, or minor loops, from major hysteresis curves, inverting these curves, and fitting them to the following function: 
\begin{equation}
	\Delta H (M, \Delta M) = I^{-1}\left(\frac{1-M}{2}\right)-I^{-1}\left(\frac{1-(M+\Delta M)}{2}\right).
	\label{fiteqn}
\end{equation}
Here, $I^{-1}$ is given by:
\begin{equation}
	I^{-1}\left(\frac{1-M}{2}\right) = -\sqrt{2} \sigma_\mathrm{I} \frac{\mathrm{erf}^{-1}(M)}{1+\gamma M}-\frac{w}{2}\frac{\tan (\frac{\pi}{2}M)}{1+\beta M}.
\end{equation}

The fit parameters $\sigma_\mathrm{I}$ and $w$ describe the intrinsic switching distribution of the array, with $\gamma$ and $\beta$ allowing for distribution asymmetry. Both Gaussian and Lorentzian forms are allowed in the model, to capture contributions that originate from both the composition of local variations (Gaussian terms originating in the central limit theorem) and possible linewidth broadening effects (Lorentzian terms). In our fits, the Lorentzian term $w$ is several orders of magnitude smaller than the Gaussian contribution $\sigma_\mathrm{I}$, which underlines the origin of the switching field distribution in the composition of multiple local variations ($\mathrm{i.e.}\:$disorder) in individual island structure.

The values of $\sigma_\mathrm{I}$ from these fits agree well with the values of $\sigma_d$ obtained from fitting to $\mathrm{Eqn.}\:$\ref{SigmaEquation}. These values are consistent within a margin of error across all lattice types and inter-island spacings, which supports the hypothesis that the disorder contribution arises from local variations in individual island properties and does not contain significant contributions due to variations arising from lattice geometry.

Equations \ref{SigmaEquation} and \ref{fiteqn} give consistent values for the width due to physical island properties, but it remains to be verified that $\mathrm{Eqn.}\:$\ref{SigmaEquation} accurately models dipolar contributions. We have measured the physical parameters that describe $B(r)$: $M_s=375\times 10^3$ A/m from SQUID measurements, $V=\pi r_0^2h$ and for our islands $r_0 =225$ nm and $h=10.4$ nm,and $K$ for each lattice is listed previously. Using these parameters and the value of $a$ from fitting $\sigma$, we can calculate the value of the proportionality constant $\alpha$.\par

\begin{table}[h]
\begin{center}
\begin{tabular}{c|c|c|c|c}
\hline\hline
\mc{}& \mc{Hex} & \mc{Kag} & \mc{Squ} & \mc{Tri} \\
\hline 
\B
 \textbf{$\sigma$} & $23.66\pm 1.96$ & $23.63\pm 1.92$  & $25.38\pm 2.20$ & $31.06\pm 0.56$\T \\
\B
\textbf{$\sigma_{\mathrm{I}}$}& $18.24\pm 1.44$ & $13.55\pm 4.32$  & $15.52\pm 3.97$ & $15.39\pm 0.53$ \T\\
\B
 \textbf{$\alpha_\mathrm{F}$} & $0.84\pm0.21$ & $0.69\pm0.17$  & $0.78\pm0.18$& $0.97\pm0.05$\T\\
\hline\hline
\end{tabular}

\end{center}
	\caption{Values of $\sigma$ and $\alpha$ as a function of lattice geometry for the second fabricated sample with 600 nm inter-island spacing. $\sigma$ and $\sigma_{\mathrm{I}}$ represent the distribution widths for the full switching field distribution and the calculated intrinsic distribution, and $\sigma_\mathrm{I}$ is interchangable with $\sigma_d$.  $\alpha_\mathrm{F}$ are values of $\alpha$ with fixed values of the parameter $\sigma_d = 15.7\pm 0.98$.}
	\label{SigmaTable}
\end{table}

The resulting values of $\alpha$ are shown in Figure \ref{SigmaData}. Values from the initial fit with variable $\sigma_d$ are denoted $\alpha_V$ and values from the second fit treating $\sigma_d$ as a fixed global parameter are denoted $\alpha_F$. They are constant within the margin of error. 
This suggests that all significant differences in the switching field distribution due to variations in lattice geometry are adequately accounted for by the effective coordination number $K$. Physically, this supports the idea that, at least in this range of interaction strengths, the dominant cause of distribution broadening is the overall magnitude of the field experienced by an island from its neighbors and is unrelated to precise details of the geometric arrangement. 
Averaging across the different geometries we find $\langle\alpha_{V}\rangle = 0.92 \pm 0.04$  and $\langle\alpha_{F}\rangle = 0.92 \pm 0.02$.



$\mathrm{Eqn.}\:$\ref{SigmaEquation} describes both the intrinsic distribution and the broadening due to dipolar interactions accurately and works well as a systematic approach to studying arrays as a function of inter-island spacing. To verify that $\alpha$ is a general quantity, we applied our analysis to a different set of fabricated arrays, considering now only the most strongly interacting arrays from that chip. These arrays have different parameters, specifically $M_s=346 \times 10^3$ A/m and  $r_0=200$ nm. The smallest inter-island spacing is $L=600$ nm. Because these arrays were fabricated at a different time with different parameters, we expect the value of $\sigma_d$ for these lattices to differ from the previous set. However, since we verified with the previous samples that $\sigma_\mathrm{I}$ gives a reasonable approximation for $\sigma_d$, we can find this parameter using the minor loops method described previously. Again, we find the average value of $\sigma_d$ from all geometries and treat it as a global variable. The values of $\sigma$ for these different arrays are shown in Table \ref{SigmaTable}, along with the value of $\alpha$ calculated using Eqn \ref{SigmaEquation} with the new parameters and fixed inter-island spacing $L$. It is not surprising that the error in this measurement is larger, since we are including information from different lattice geometries at a fixed spacing, instead of fitting each lattice type across a range of spacings.
To find the average value of $\alpha_F$, taking into account the different errors, a weighted average is used. The values are weighted by the variance, 
$\langle\alpha_F\rangle=\sum_i\alpha_{Fi}\sigma_i^{-2}/\sum_i\sigma_i^{-2}$.
 This value, $\langle\alpha_F\rangle = 0.93\pm0.09$, agrees well with that calculated for the other chip. 
 
In summary, we have demonstrated that using diffraction-limited MOKE imaging combined with appropriate image processing techniques, we can reliably find the switching fields of individual islands within a large array of perpendicular nanomagnets. This information allows us to directly measure the switching field distribution, which we can then analytically interpret to isolate the contributions from dipolar interactions and disorder due to individual island properties. By confirming the efficacy of our numerical analysis using the refined hysteresis loops from individual island switching fields, we have verified a global analysis method that allows for quick characterizion of the strength of disorder. 
Quantifying this disorder strength is important for understanding the quality of the samples. Interactions can be designed in an idealized way during the fabrication process, but the actual disorder present is an important factor in how the islands will physically behave and how close the arrays can be brought to the ground state, and this disorder can only be determined post-fabrication. Moreover, accessing individual island information from a quasi-dynamic measurement with an \textit{in situ} applied field opens the door to further studies of dynamics and correlations that could lead to a much richer understanding of the behavior of systems governed by dipolar interactions. 

This project was funded by the US Department of Energy, Office of Basic Energy Sciences, Materials Sciences and Engineering Division under Grant No. DE-SC0010778.


\begin{thebibliography}{17}
\makeatletter
\providecommand \@ifxundefined [1]{%
 \@ifx{#1\undefined}
}%
\providecommand \@ifnum [1]{%
 \ifnum #1\expandafter \@firstoftwo
 \else \expandafter \@secondoftwo
 \fi
}%
\providecommand \@ifx [1]{%
 \ifx #1\expandafter \@firstoftwo
 \else \expandafter \@secondoftwo
 \fi
}%
\providecommand \natexlab [1]{#1}%
\providecommand \enquote  [1]{``#1''}%
\providecommand \bibnamefont  [1]{#1}%
\providecommand \bibfnamefont [1]{#1}%
\providecommand \citenamefont [1]{#1}%
\providecommand \href@noop [0]{\@secondoftwo}%
\providecommand \href [0]{\begingroup \@sanitize@url \@href}%
\providecommand \@href[1]{\@@startlink{#1}\@@href}%
\providecommand \@@href[1]{\endgroup#1\@@endlink}%
\providecommand \@sanitize@url [0]{\catcode `\\12\catcode `\$12\catcode
  `\&12\catcode `\#12\catcode `\^12\catcode `\_12\catcode `\%12\relax}%
\providecommand \@@startlink[1]{}%
\providecommand \@@endlink[0]{}%
\providecommand \url  [0]{\begingroup\@sanitize@url \@url }%
\providecommand \@url [1]{\endgroup\@href {#1}{\urlprefix }}%
\providecommand \urlprefix  [0]{URL }%
\providecommand \Eprint [0]{\href }%
\providecommand \doibase [0]{http://dx.doi.org/}%
\providecommand \selectlanguage [0]{\@gobble}%
\providecommand \bibinfo  [0]{\@secondoftwo}%
\providecommand \bibfield  [0]{\@secondoftwo}%
\providecommand \translation [1]{[#1]}%
\providecommand \BibitemOpen [0]{}%
\providecommand \bibitemStop [0]{}%
\providecommand \bibitemNoStop [0]{.\EOS\space}%
\providecommand \EOS [0]{\spacefactor3000\relax}%
\providecommand \BibitemShut  [1]{\csname bibitem#1\endcsname}%
\let\auto@bib@innerbib\@empty
\bibitem [{\citenamefont {Wang}\ \emph {et~al.}(2006)\citenamefont {Wang},
  \citenamefont {Nisoli}, \citenamefont {Freitas}, \citenamefont {Li},
  \citenamefont {McConville}, \citenamefont {Cooley}, \citenamefont {Lund},
  \citenamefont {Samarth}, \citenamefont {Leighton}, \citenamefont {Crespi},\
  and\ \citenamefont {Schiffer}}]{Wang2006}%
  \BibitemOpen
  \bibfield  {author} {\bibinfo {author} {\bibfnamefont {R.~F.}\ \bibnamefont
  {Wang}}, \bibinfo {author} {\bibfnamefont {C.}~\bibnamefont {Nisoli}},
  \bibinfo {author} {\bibfnamefont {R.~S.}\ \bibnamefont {Freitas}}, \bibinfo
  {author} {\bibfnamefont {J.}~\bibnamefont {Li}}, \bibinfo {author}
  {\bibfnamefont {W.}~\bibnamefont {McConville}}, \bibinfo {author}
  {\bibfnamefont {B.~J.}\ \bibnamefont {Cooley}}, \bibinfo {author}
  {\bibfnamefont {M.~S.}\ \bibnamefont {Lund}}, \bibinfo {author}
  {\bibfnamefont {N.}~\bibnamefont {Samarth}}, \bibinfo {author} {\bibfnamefont
  {C.}~\bibnamefont {Leighton}}, \bibinfo {author} {\bibfnamefont {V.~H.}\
  \bibnamefont {Crespi}}, \ and\ \bibinfo {author} {\bibfnamefont
  {P.}~\bibnamefont {Schiffer}},\ }\href {\doibase 10.1038/nature04447}
  {\bibfield  {journal} {\bibinfo  {journal} {Nature}\ }\textbf {\bibinfo
  {volume} {439}},\ \bibinfo {pages} {303} (\bibinfo {year}
  {2006})}\BibitemShut {NoStop}%
\bibitem [{\citenamefont {Qi}\ \emph {et~al.}(2008)\citenamefont {Qi},
  \citenamefont {Brintlinger},\ and\ \citenamefont {Cumings}}]{Qi2008}%
  \BibitemOpen
  \bibfield  {author} {\bibinfo {author} {\bibfnamefont {Y.}~\bibnamefont
  {Qi}}, \bibinfo {author} {\bibfnamefont {T.}~\bibnamefont {Brintlinger}}, \
  and\ \bibinfo {author} {\bibfnamefont {J.}~\bibnamefont {Cumings}},\ }\href
  {\doibase 10.1103/PhysRevB.77.094418} {\bibfield  {journal} {\bibinfo
  {journal} {Phys. Rev. B}\
  }\textbf {\bibinfo {volume} {77}},\ \bibinfo {pages} {1} (\bibinfo {year}
  {2008})}
  \BibitemShut {NoStop}%
\bibitem [{\citenamefont {Nisoli}\ \emph {et~al.}(2013)\citenamefont {Nisoli},
  \citenamefont {Moessner},\ and\ \citenamefont {Schiffer}}]{Nisoli2013}%
  \BibitemOpen
  \bibfield  {author} {\bibinfo {author} {\bibfnamefont {C.}~\bibnamefont
  {Nisoli}}, \bibinfo {author} {\bibfnamefont {R.}~\bibnamefont {Moessner}}, \
  and\ \bibinfo {author} {\bibfnamefont {P.}~\bibnamefont {Schiffer}},\ }\href
  {\doibase 10.1103/RevModPhys.85.1473} {\bibfield  {journal} {\bibinfo
  {journal} {Rev. Mod. Phys.}\ }\textbf {\bibinfo {volume} {85}},\
  \bibinfo {pages} {1473} (\bibinfo {year} {2013})} \BibitemShut {NoStop}%
\bibitem [{\citenamefont {Ke}\ \emph {et~al.}(2008)\citenamefont {Ke},
  \citenamefont {Li}, \citenamefont {Nisoli}, \citenamefont {Lammert},
  \citenamefont {McConville}, \citenamefont {Wang}, \citenamefont {Crespi},\
  and\ \citenamefont {Schiffer}}]{Ke2008b}%
  \BibitemOpen
  \bibfield  {author} {\bibinfo {author} {\bibfnamefont {X.}~\bibnamefont
  {Ke}}, \bibinfo {author} {\bibfnamefont {J.}~\bibnamefont {Li}}, \bibinfo
  {author} {\bibfnamefont {C.}~\bibnamefont {Nisoli}}, \bibinfo {author}
  {\bibfnamefont {P.~E.}\ \bibnamefont {Lammert}}, \bibinfo {author}
  {\bibfnamefont {W.}~\bibnamefont {McConville}}, \bibinfo {author}
  {\bibfnamefont {R.~F.}\ \bibnamefont {Wang}}, \bibinfo {author}
  {\bibfnamefont {V.~H.}\ \bibnamefont {Crespi}}, \ and\ \bibinfo {author}
  {\bibfnamefont {P.}~\bibnamefont {Schiffer}},\ }\href {\doibase
  10.1103/PhysRevLett.101.037205} {\bibfield  {journal} {\bibinfo  {journal}
  {Phys. Rev. Lett.}\ }\textbf {\bibinfo {volume} {101}},\ \bibinfo
  {pages} {1} (\bibinfo {year} {2008})}
   \BibitemShut {NoStop}%
\bibitem [{\citenamefont {Morgan}\ \emph {et~al.}(2011)\citenamefont {Morgan},
  \citenamefont {Stein}, \citenamefont {Langridge},\ and\ \citenamefont
  {Marrows}}]{Morgan2011}%
  \BibitemOpen
  \bibfield  {author} {\bibinfo {author} {\bibfnamefont {J.~P.}\ \bibnamefont
  {Morgan}}, \bibinfo {author} {\bibfnamefont {A.}~\bibnamefont {Stein}},
  \bibinfo {author} {\bibfnamefont {S.}~\bibnamefont {Langridge}}, \ and\
  \bibinfo {author} {\bibfnamefont {C.~H.}\ \bibnamefont {Marrows}},\ }\href
  {\doibase 10.1088/1367-2630/13/10/105002} {\bibfield  {journal} {\bibinfo
  {journal} {New J. Phys.}\ }\textbf {\bibinfo {volume} {13}}, {\bibinfo {pages} {105002}}
  (\bibinfo {year} {2011})}\BibitemShut
  {NoStop}%
\bibitem [{\citenamefont {Morgan}\ \emph {et~al.}(2013)\citenamefont {Morgan},
  \citenamefont {Bellew}, \citenamefont {Stein}, \citenamefont {Langridge},\
  and\ \citenamefont {Marrows}}]{Morgan2013}%
  \BibitemOpen
  \bibfield  {author} {\bibinfo {author} {\bibfnamefont {J.~P.}\ \bibnamefont
  {Morgan}}, \bibinfo {author} {\bibfnamefont {A.}~\bibnamefont {Bellew}},
  \bibinfo {author} {\bibfnamefont {A.}~\bibnamefont {Stein}}, \bibinfo
  {author} {\bibfnamefont {S.}~\bibnamefont {Langridge}}, \ and\ \bibinfo
  {author} {\bibfnamefont {C.~H.}\ \bibnamefont {Marrows}},\ }\href {\doibase
  10.3389/fphy.2013.00028} {\bibfield  {journal} {\bibinfo  {journal}
  {Frontiers in Condensed Matter Physics}\ }\textbf {\bibinfo {volume} {1}},\
  \bibinfo {pages} {28} (\bibinfo {year} {2013})}\BibitemShut {NoStop}%
\bibitem [{\citenamefont {Kapaklis}\ \emph {et~al.}(2014)\citenamefont
  {Kapaklis}, \citenamefont {Arnalds}, \citenamefont {Farhan}, \citenamefont
  {Chopdekar}, \citenamefont {Balan}, \citenamefont {Scholl}, \citenamefont
  {Heyderman},\ and\ \citenamefont {Hj{\"{o}}rvarsson}}]{Kapaklis2014}%
  \BibitemOpen
  \bibfield  {author} {\bibinfo {author} {\bibfnamefont {V.}~\bibnamefont
  {Kapaklis}}, \bibinfo {author} {\bibfnamefont {U.~B.}\ \bibnamefont
  {Arnalds}}, \bibinfo {author} {\bibfnamefont {A.}~\bibnamefont {Farhan}},
  \bibinfo {author} {\bibfnamefont {R.~V.}\ \bibnamefont {Chopdekar}}, \bibinfo
  {author} {\bibfnamefont {A.}~\bibnamefont {Balan}}, \bibinfo {author}
  {\bibfnamefont {A.}~\bibnamefont {Scholl}}, \bibinfo {author} {\bibfnamefont
  {L.~J.}\ \bibnamefont {Heyderman}}, \ and\ \bibinfo {author} {\bibfnamefont
  {B.}~\bibnamefont {Hj{\"{o}}rvarsson}},\ }\href {\doibase
  10.1038/nnano.2014.104} {\bibfield  {journal} {\bibinfo  {journal} {Nature
  Nano.}\ }\textbf {\bibinfo {volume} {9}},\ \bibinfo {pages} {514}
  (\bibinfo {year} {2014})}\BibitemShut {NoStop}%
\bibitem [{\citenamefont {Ladak}\ \emph {et~al.}(2011)\citenamefont {Ladak},
  \citenamefont {Read}, \citenamefont {Branford},\ and\ \citenamefont
  {Cohen}}]{Ladak2011}%
  \BibitemOpen
  \bibfield  {author} {\bibinfo {author} {\bibfnamefont {S.}~\bibnamefont
  {Ladak}}, \bibinfo {author} {\bibfnamefont {D.~E.}\ \bibnamefont {Read}},
  \bibinfo {author} {\bibfnamefont {W.~R.}\ \bibnamefont {Branford}}, \ and\
  \bibinfo {author} {\bibfnamefont {L.~F.}\ \bibnamefont {Cohen}},\ }\href
  {\doibase 10.1088/1367-2630/13/6/063032} {\bibfield  {journal} {\bibinfo
  {journal} {New J. Phys.}\ }\textbf {\bibinfo {volume} {13}},\
  \bibinfo {pages} {359} (\bibinfo {year} {2011})}\BibitemShut {NoStop}%
\bibitem [{\citenamefont {Mengotti}\ \emph {et~al.}(2011)\citenamefont
  {Mengotti}, \citenamefont {Heyderman}, \citenamefont {Rodr{\'{\i}}guez},
  \citenamefont {Nolting}, \citenamefont {H{\"{u}}gli},\ and\ \citenamefont
  {Braun}}]{Mengotti2011}%
  \BibitemOpen
  \bibfield  {author} {\bibinfo {author} {\bibfnamefont {E.}~\bibnamefont
  {Mengotti}}, \bibinfo {author} {\bibfnamefont {L.~J.}\ \bibnamefont
  {Heyderman}}, \bibinfo {author} {\bibfnamefont {A.~F.}\ \bibnamefont
  {Rodr{\'{\i}}guez}}, \bibinfo {author} {\bibfnamefont {F.}~\bibnamefont
  {Nolting}}, \bibinfo {author} {\bibfnamefont {R.~V.}\ \bibnamefont
  {H{\"{u}}gli}}, \ and\ \bibinfo {author} {\bibfnamefont {H.-B.}\ \bibnamefont
  {Braun}},\ }\href {\doibase 10.1038/nphys1794} {\bibfield  {journal}
  {\bibinfo  {journal} {Nature Phys.}\ }\textbf {\bibinfo {volume} {7}},\
  \bibinfo {pages} {68} (\bibinfo {year} {2011})}\BibitemShut {NoStop}%
\bibitem [{\citenamefont {Budrikis}(2014)}]{Budrikis2014}%
  \BibitemOpen
  \bibfield  {author} {\bibinfo {author} {\bibfnamefont {Z.}~\bibnamefont
  {Budrikis}},\ }in\ \href {\doibase {10.1016/B978-0-12-800175-2.00002-9}}
  {\emph {\bibinfo {booktitle} {{Solid State Physics, Vol. 65}}}},\ \bibinfo
  {editor} {edited by\ \bibinfo {editor} {\bibnamefont {{R. E.Camley and R. L. Stamps}}}}\ (\bibinfo {year} {{2014}})\ pp.\ \bibinfo {pages}
  {{109--236}}\BibitemShut {NoStop}%
\bibitem [{\citenamefont {Mengotti}\ \emph {et~al.}(2009)\citenamefont
  {Mengotti}, \citenamefont {Heyderman}, \citenamefont {Bisig}, \citenamefont
  {{Fraile Rodr{\'{\i}}guez}}, \citenamefont {{Le Guyader}}, \citenamefont
  {Nolting},\ and\ \citenamefont {Braun}}]{Mengotti2009}%
  \BibitemOpen
  \bibfield  {author} {\bibinfo {author} {\bibfnamefont {E.}~\bibnamefont
  {Mengotti}}, \bibinfo {author} {\bibfnamefont {L.~J.}\ \bibnamefont
  {Heyderman}}, \bibinfo {author} {\bibfnamefont {A.}~\bibnamefont {Bisig}},
  \bibinfo {author} {\bibfnamefont {A.~F.}~\bibnamefont {{Rodr{\'{\i}}guez}}}, \bibinfo {author} {\bibfnamefont {L.}~\bibnamefont {{Le
  Guyader}}}, \bibinfo {author} {\bibfnamefont {F.}~\bibnamefont {Nolting}}, \
  and\ \bibinfo {author} {\bibfnamefont {H.~B.}\ \bibnamefont {Braun}},\ }\href
  {\doibase 10.1063/1.3133202} {\bibfield  {journal} {\bibinfo  {journal}
  {J. Appl. Phys.}\ }\textbf {\bibinfo {volume} {105}}, \
  \bibinfo {pages} {113113} (\bibinfo
  {year} {2009})}\BibitemShut {NoStop}%
\bibitem [{\citenamefont {Zhang}\ \emph {et~al.}(2012)\citenamefont {Zhang},
  \citenamefont {Li}, \citenamefont {Gilbert}, \citenamefont {Bartell},
  \citenamefont {Erickson}, \citenamefont {Pan}, \citenamefont {Lammert},
  \citenamefont {Nisoli}, \citenamefont {Kohli}, \citenamefont {Misra},
  \citenamefont {Crespi}, \citenamefont {Samarth}, \citenamefont {Leighton},\
  and\ \citenamefont {Schiffer}}]{Zhang2012a}%
  \BibitemOpen
  \bibfield  {author} {\bibinfo {author} {\bibfnamefont {S.}~\bibnamefont
  {Zhang}}, \bibinfo {author} {\bibfnamefont {J.}~\bibnamefont {Li}}, \bibinfo
  {author} {\bibfnamefont {I.}~\bibnamefont {Gilbert}}, \bibinfo {author}
  {\bibfnamefont {J.}~\bibnamefont {Bartell}}, \bibinfo {author} {\bibfnamefont
  {M.~J.}\ \bibnamefont {Erickson}}, \bibinfo {author} {\bibfnamefont
  {Y.}~\bibnamefont {Pan}}, \bibinfo {author} {\bibfnamefont {P.~E.}\
  \bibnamefont {Lammert}}, \bibinfo {author} {\bibfnamefont {C.}~\bibnamefont
  {Nisoli}}, \bibinfo {author} {\bibfnamefont {K.~K.}\ \bibnamefont {Kohli}},
  \bibinfo {author} {\bibfnamefont {R.}~\bibnamefont {Misra}}, \bibinfo
  {author} {\bibfnamefont {V.~H.}\ \bibnamefont {Crespi}}, \bibinfo {author}
  {\bibfnamefont {N.}~\bibnamefont {Samarth}}, \bibinfo {author} {\bibfnamefont
  {C.}~\bibnamefont {Leighton}}, \ and\ \bibinfo {author} {\bibfnamefont
  {P.}~\bibnamefont {Schiffer}},\ }\href {\doibase
  10.1103/PhysRevLett.109.087201} {\bibfield  {journal} {\bibinfo  {journal}
  {Phys. Rev. Lett.}\ }\textbf {\bibinfo {volume} {109}}, \
  \bibinfo {pages} {087201} (\bibinfo
  {year} {2012}).}\BibitemShut {NoStop}%
\bibitem [{\citenamefont {Li}\ \emph {et~al.}(2011)\citenamefont {Li},
  \citenamefont {Chen}, \citenamefont {Huang}, \citenamefont {Xue},\ and\
  \citenamefont {Ding}}]{Li2011}%
  \BibitemOpen
  \bibfield  {author} {\bibinfo {author} {\bibfnamefont {W.~M.}\ \bibnamefont
  {Li}}, \bibinfo {author} {\bibfnamefont {Y.~J.}\ \bibnamefont {Chen}},
  \bibinfo {author} {\bibfnamefont {T.~L.}\ \bibnamefont {Huang}}, \bibinfo
  {author} {\bibfnamefont {J.~M.}\ \bibnamefont {Xue}}, \ and\ \bibinfo
  {author} {\bibfnamefont {J.}~\bibnamefont {Ding}},\ }\href {\doibase
  10.1063/1.3563069} {\bibfield  {journal} {\bibinfo  {journal} {J. Appl. Phys.}\ }\textbf {\bibinfo {volume} {109}},\ \bibinfo {pages} {1}
  (\bibinfo {year} {2011})}\BibitemShut {NoStop}%
\bibitem [{\citenamefont {Cormier}\ \emph {et~al.}(2008)\citenamefont
  {Cormier}, \citenamefont {Ferŕ}, \citenamefont {Mougin}, \citenamefont
  {Crom{\`{\i}}res},\ and\ \citenamefont {Klein}}]{Cormier2008}%
  \BibitemOpen
  \bibfield  {author} {\bibinfo {author} {\bibfnamefont {M.}~\bibnamefont
  {Cormier}}, \bibinfo {author} {\bibfnamefont {J.}~\bibnamefont {Ferŕ}},
  \bibinfo {author} {\bibfnamefont {a.}~\bibnamefont {Mougin}}, \bibinfo
  {author} {\bibfnamefont {J.~P.}\ \bibnamefont {Crom{\`{\i}}res}}, \ and\
  \bibinfo {author} {\bibfnamefont {V.}~\bibnamefont {Klein}},\ }\href
  {\doibase 10.1063/1.2890839} {\bibfield  {journal} {\bibinfo  {journal}
  {Rev. Sci. Instrum.}\ }\textbf {\bibinfo {volume} {79}},\
  \bibinfo {pages} {1} (\bibinfo {year} {2008})}\BibitemShut {NoStop}%
\bibitem [{\citenamefont {Lin}\ \emph {et~al.}(1991)\citenamefont {Lin},
  \citenamefont {Gorman}, \citenamefont {Lee}, \citenamefont {Farrow},
  \citenamefont {Marinero}, \citenamefont {Do}, \citenamefont {Notarys},\ and\
  \citenamefont {Chien}}]{Lin1991}%
  \BibitemOpen
  \bibfield  {author} {\bibinfo {author} {\bibfnamefont {C.-J.}\ \bibnamefont
  {Lin}}, \bibinfo {author} {\bibfnamefont {G.}~\bibnamefont {Gorman}},
  \bibinfo {author} {\bibfnamefont {C.}~\bibnamefont {Lee}}, \bibinfo {author}
  {\bibfnamefont {R.}~\bibnamefont {Farrow}}, \bibinfo {author} {\bibfnamefont
  {E.}~\bibnamefont {Marinero}}, \bibinfo {author} {\bibfnamefont
  {H.}~\bibnamefont {Do}}, \bibinfo {author} {\bibfnamefont {H.}~\bibnamefont
  {Notarys}}, \ and\ \bibinfo {author} {\bibfnamefont {C.}~\bibnamefont
  {Chien}},\ }\href {\doibase 10.1016/0304-8853(91)90329-9} {\bibfield
  {journal} {\bibinfo  {journal} {J. Magn. Magn. Mater.}\
  }\textbf {\bibinfo {volume} {93}},\ \bibinfo {pages} {194} (\bibinfo {year}
  {1991})}\BibitemShut {NoStop}%
\bibitem [{\citenamefont {Gerrits}\ \emph {et~al.}(2002)\citenamefont
  {Gerrits}, \citenamefont {van~den Berg}, \citenamefont {Hohlfeld},
  \citenamefont {B{\"{a}}r},\ and\ \citenamefont {Rasing}}]{Gerrits2002}%
  \BibitemOpen
  \bibfield  {author} {\bibinfo {author} {\bibfnamefont {Th.}\ \bibnamefont
  {Gerrits}}, \bibinfo {author} {\bibfnamefont {H.~A.~M.}\ \bibnamefont
  {van~den Berg}}, \bibinfo {author} {\bibfnamefont {J.}~\bibnamefont
  {Hohlfeld}}, \bibinfo {author} {\bibfnamefont {L.}~\bibnamefont {B{\"{a}}r}},
  \ and\ \bibinfo {author} {\bibfnamefont {Th.}\ \bibnamefont {Rasing}},\
  }\href {\doibase 10.1038/nature00905} {\bibfield  {journal} {\bibinfo
  {journal} {Nature}\ }\textbf {\bibinfo {volume} {418}},\ \bibinfo {pages}
  {509} (\bibinfo {year} {2002})}\BibitemShut {NoStop}%
\bibitem [{\citenamefont {Berger}\ \emph {et~al.}(2005)\citenamefont {Berger},
  \citenamefont {Xu}, \citenamefont {Lengsfield}, \citenamefont {Ikeda},\ and\
  \citenamefont {Fullerton}}]{Berger2005}%
  \BibitemOpen
  \bibfield  {author} {\bibinfo {author} {\bibfnamefont {A.}~\bibnamefont
  {Berger}}, \bibinfo {author} {\bibfnamefont {Y.}~\bibnamefont {Xu}}, \bibinfo
  {author} {\bibfnamefont {B.}~\bibnamefont {Lengsfield}}, \bibinfo {author}
  {\bibfnamefont {Y.}~\bibnamefont {Ikeda}}, \ and\ \bibinfo {author}
  {\bibfnamefont {E.~E.}\ \bibnamefont {Fullerton}},\ }\href {\doibase
  10.1109/TMAG.2005.855285} {\bibfield  {journal} {\bibinfo  {journal} {IEEE
  Trans. Mag.}\ }\textbf {\bibinfo {volume} {41}},\ \bibinfo
  {pages} {3178} (\bibinfo {year} {2005})}
  \BibitemShut {NoStop}%
\end{thebibliography}

\newpage 

\newpage

\end{document}